\documentclass[a4paper,11pt]{article}
\usepackage{pos}
\usepackage{graphicx}

\title{White Dwarf Pulsars: an emerging class of compact binaries}

\author*[a,b,c]{David A.H. Buckley}

\affiliation[a]{South African Astronomical Observatory, PO Box 9, Observatory Road, Observatory 7935, Cape Town, South Africa}

\affiliation[b]{Department of Astronomy, University of Cape Town, Private Bag X3, Rondebosch 7701, South Africa}

\affiliation[c]{Department of Physics, University of the Free State, PO Box 339, Bloemfontein 9300, South Africa}

\emailAdd{DAH.Buckley@saao.nrf.ac.za}

\abstract{I summarise the expanding class of white dwarf pulsars for which the prototype, AR Scorpii, was only discovered in 2016. At the time of this meeting, two further systems had been found (eRASSU J191213.9-441044 and  SDSS J230641.47+244055.8). These are thought to be detached (or nearly detached) white-dwarf--red-dwarf binaries, where the white dwarf's magnetic field interacts with the M-star's corona, resulting in the capture and acceleration of electrons to relativistic velocities in the magnetosphere. This gives rise to non-thermal synchrotron emission across the electromagnetic spectrum. The first discovered, and brightest, system, AR Sco (which is the focus of this paper), shows a white dwarf spin-down consistent with magnetic dipole emission in a strongly magnetic white dwarf. The polarisation properties of the two systems for which polarimetry has been obtained (AR Sco and eRASSU J191213.9-441044) show spin-modulated polarisation, primarily linear, which remarkably reaches 40\% in AR Sco. Furthermore, the Stokes Q and U parameters vary in the same manner as the optical polarisation of the Crab pulsar, consistent with a rotating magnetic dipole. These systems are intriguing in that each one seems to show different behaviours (e.g. pulse profiles and light-curve properties), presumably mainly due to different viewing aspect angles. It has recently been postulated that these systems represent a short-lived phase of binary evolution linking asynchronously rotating intermediate polars and synchronous polars. }

\FullConference{
}

\begin{document}
\maketitle

\section{Introduction}

Rotationally powered pulsars have traditionally been associated with neutron
stars, whose rapid rotation and intense magnetic fields sustain
magnetospheres capable of accelerating particles to relativistic energies.
Although magnetic white dwarfs possess substantial magnetic fields and
rotational energy reservoirs, they were not generally expected to exhibit
analogous pulsar-like behaviour.

This picture changed dramatically with the discovery of the nature of AR Sco \cite{marsh2016,buckley2017}. 
The system consists of a rapidly rotating ($\sim$2 min period) magnetic white
dwarf in a 3.56\,h binary with an M-dwarf companion. Unlike classical
magnetic cataclysmic variables, its luminosity is powered predominantly by
the loss of rotational kinetic energy rather than by accretion. Coherent
pulsations are observed across the electromagnetic spectrum, from radio to
X-rays, making AR Sco the first unambiguous example of a white dwarf pulsar.

Subsequent observational campaigns have transformed our understanding of
this remarkable object. High-speed optical photometry, spectroscopy,
polarimetry, radio observations with MeerKAT, and X-ray observations with
NICER and XMM-Newton have provided an increasingly coherent picture of a
magnetically dominated system in which synchrotron radiation from
relativistic particles is modulated by the white dwarf spin and its
interaction with the companion star.

The recent discoveries of eRASSU~J191213.9$-$441044 \cite{pelisoli2023,Schwope2023} and
SDSS~J230641.47+244055.8 \cite{castro2025} demonstrate that AR Sco is not unique. Instead,
these systems appear to define an emerging class of compact interacting
binaries in which rotational energy loss powers non-thermal emission. While they
share many of the defining characteristics of AR Sco, important differences
between them provide new insights into the range of physical conditions under
which white dwarf pulsars can exist, as well as the effects of viewing geometry (i.e. the viewing angle and the offset of the magnetic axis from the spin axis).

This review summarises the observational properties of the currently known
white dwarf pulsars, particularly AR Sco itself, emphasising the multiwavelength campaigns that have
established the physical picture of these systems. We discuss their timing,
energetics, polarimetric behaviour, and emission mechanisms before
considering their evolutionary significance and the outstanding questions
that remain for this newly recognised class of compact binaries.

\section{AR Sco: the prototype white dwarf pulsar}
\label{sec:arsco}

AR Scorpii (AR Sco) is the prototype of an emerging class of white dwarf
pulsars. Its importance is not simply that it is an unusual compact binary,
but that several independent lines of evidence show that its dominant
pulsed emission is powered by the rotational energy loss of a spinning-down, highly magnetic white dwarf, consistent with magnetic-dipole emission, rather than by accretion. In this respect AR Sco provided the first clear
example of pulsar-like behaviour in a white dwarf system.

\begin{figure}
\centering
\includegraphics[width=.80\textwidth]{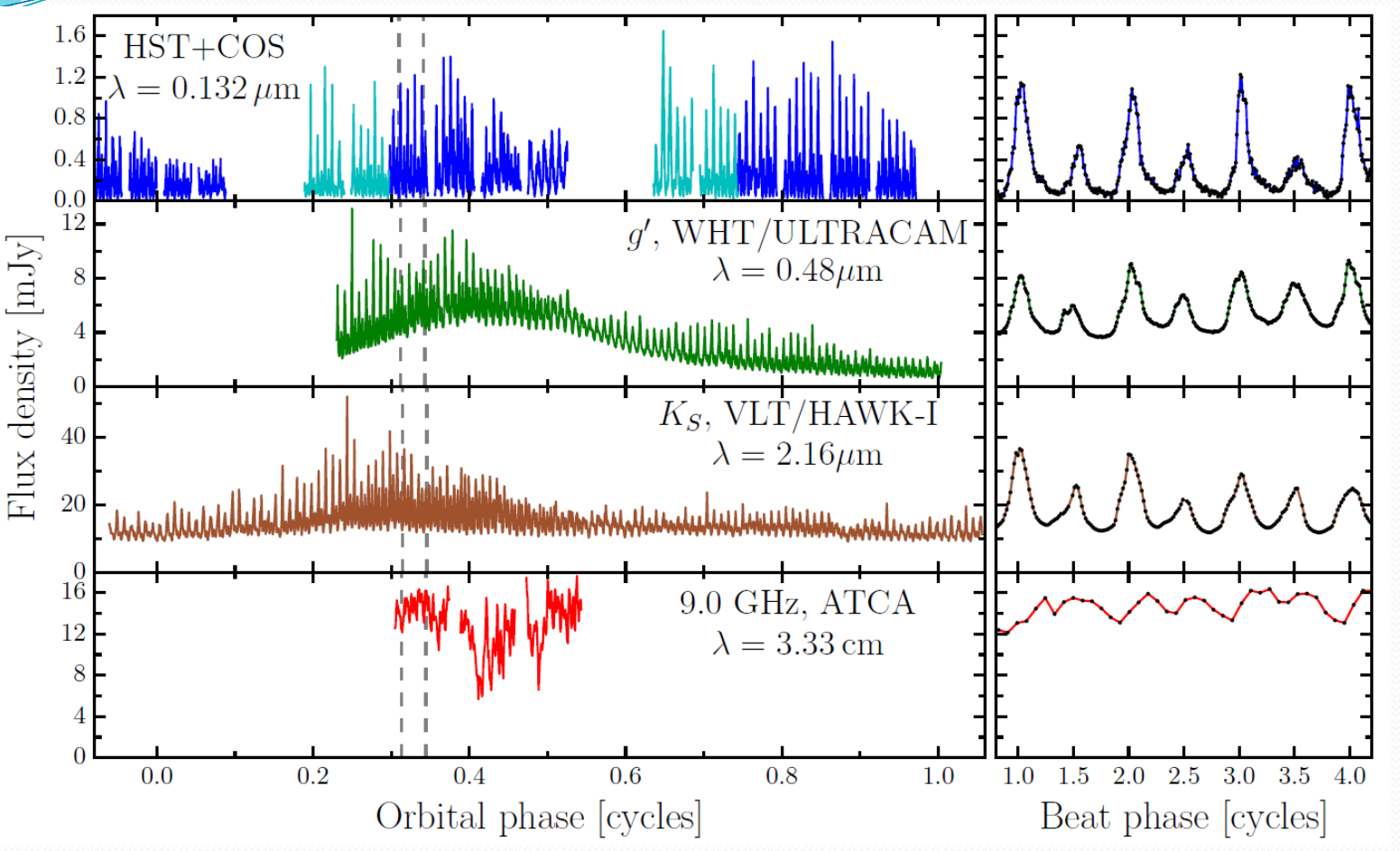}
\caption{Representative high-speed light curves of AR Sco showing
    the strong coherent pulsations across the electromagnetic spectrum (top 
    to bottom: ultraviolet, visible, infrared, radio). The right panel is a zoomed-in view in the region indicated by the two  vertical dashed lines in the left panel \cite{marsh2016}.}
\label{fig:arsco_lightcurves}
\end{figure}

\subsection{Discovery of spin modulations and basic timing properties}
\label{subsec:arsco_timing}

The first modern observations of AR Sco revealed coherent non-thermal
pulsations across the electromagnetic spectrum \cite{marsh2016} (see Figure 1). The system
contains a rapidly rotating white dwarf with a spin period of
$P_{\rm spin}=117.12$~s in a binary with an orbital period of about 3.6~h.
A second period, at $P_{\rm beat}=118.2$~s, corresponds to the beat between
the white dwarf spin and the orbital motion, and in the optical this beat
period often dominates the power spectrum. The pulse profiles are double
peaked, with a strong first harmonic, particularly at shorter wavelengths.

\begin{figure}
\centering
\includegraphics[width=.80\textwidth]{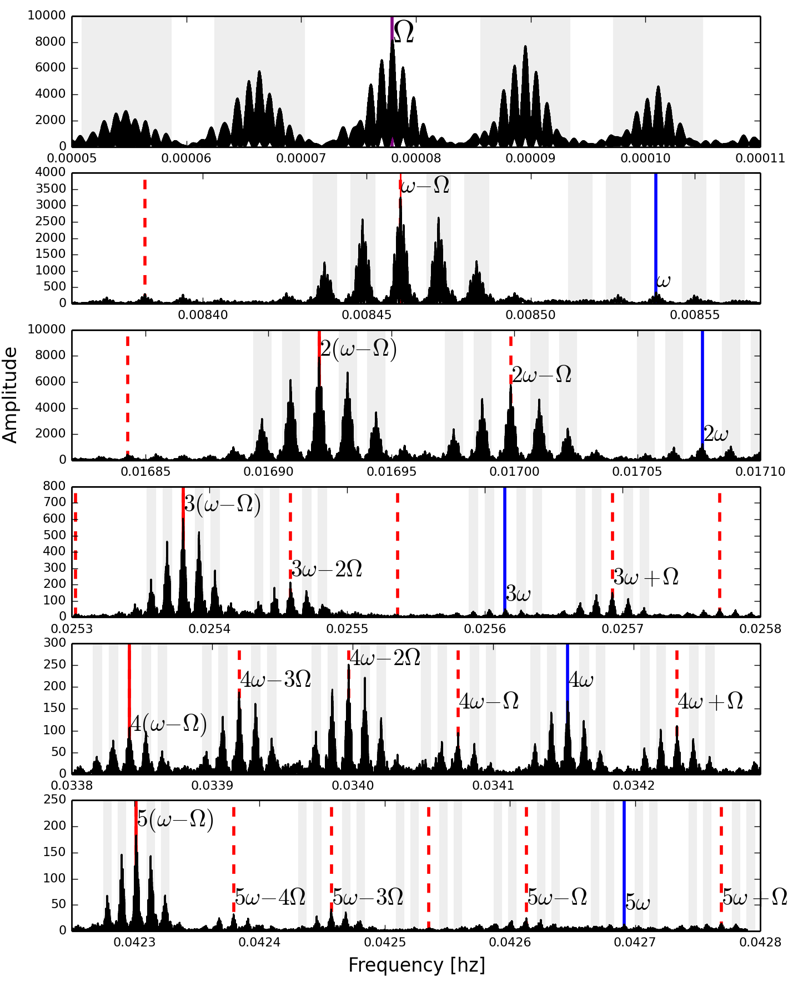}
\caption{Power spectrum of AR Sco photopolarimetry \cite{PotterBuckley2018} showing coherent signals at the white
dwarf spin frequency, the orbital beat frequency, their harmonics and aliases.}
\label{fig:arsco_powerspectrum}
\end{figure}

In Figure~\ref{fig:arsco_powerspectrum} the power spectrum of $\sim$60 nights of all-Stokes polarimetry \cite{PotterBuckley2018}, obtained over two years, is presented. All of the power is produced by the combination of two fundamental signals, namely the orbital and spin frequencies, their harmonics, the respective sideband frequencies (e.g. the beat frequency) and aliasing due to data sampling on daily, monthly and yearly timescales. The dominance of the beat modulation in the optical is one of the key
signatures of interaction between the rotating white dwarf magnetosphere
and the companion star.

The measured long-term increase in the spin period determined from sparsely sampled CRTS data over a $\sim$7-year timescale is another defining
property. The original analysis gave
$\dot{P}=3.92\times10^{-13}\,{\rm s\,s^{-1}}$ \cite{marsh2016}, implying a
synchronisation timescale of order $10^7$~yr. A later refinement gives
$\dot{P}=6.62\times10^{-13}\,{\rm s\,s^{-1}}$ \cite{pelisoli2022}. This
spin-down is central to the interpretation of AR Sco as a rotationally
powered system.

\subsection{Spectral components and non-thermal dominance}
\label{subsec:arsco_sed}

The optical spectrum of AR Sco contains contributions from the M-type
secondary star, a relatively faint white dwarf component constrained by
ultraviolet observations, and a dominant non-thermal component \cite{marsh2016} (see Figure 3). After
subtracting the stellar components, the residual optical continuum is well
described by a power law and dominates at wavelengths shorter than about
8000~\AA. The secondary star is consistent with an M5~V star, and spectral
and infrared constraints originally implied a distance of about
$116\pm16$~pc.

Spectroscopic observations also show narrow ultraviolet and optical emission
lines, together with Na~I absorption, moving in radial velocity in a manner
consistent with irradiation of the inner face of the companion. Infrared
photometry shows ellipsoidal variations, indicating that the M star is close
to filling its Roche lobe. These properties show that the companion plays an
important role in shaping the observed modulation, even though the system is
not dominated by accretion.

\begin{figure}[t]
\centering
\includegraphics[width=.90\textwidth]{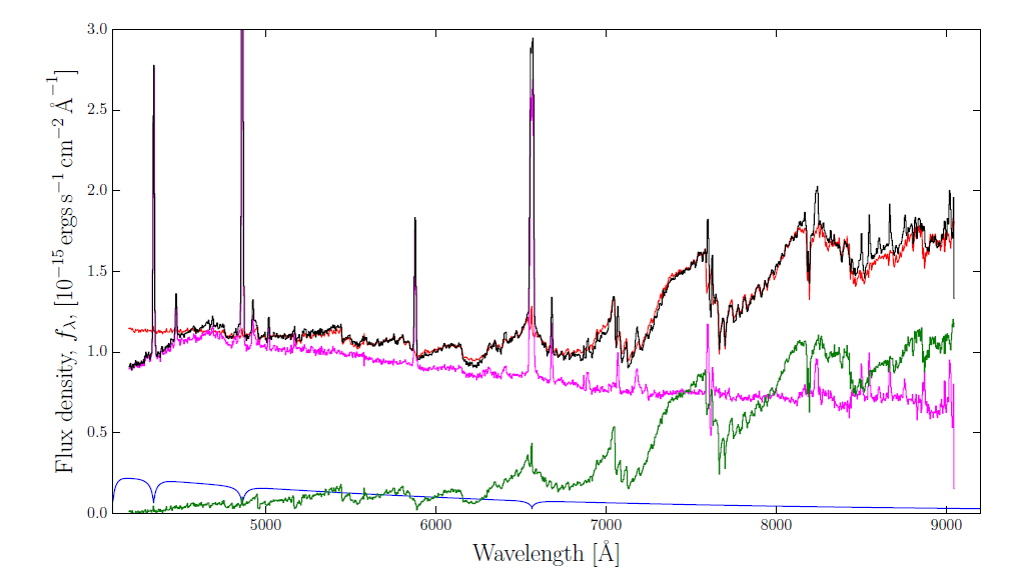}
\caption{Optical spectral decomposition of AR Sco \cite{marsh2016}. The observed spectrum
    can be separated into an M-dwarf contribution (green), a constrained white dwarf
    component (blue), and a dominant non-thermal continuum (magenta).}
\label{fig:arsco_spectrum}
\end{figure}

More recent spectral-energy-distribution modelling (e.g. \cite{Geng2016,Takata2018})
supports a picture in
which two synchrotron components are present. A low-frequency component below
$\sim10^{13}$~Hz, extending from radio to infrared wavelengths, has been
associated with magnetically confined or pumped coronal-loop structures.
A higher-frequency component, extending from the infrared through optical,
ultraviolet, and X-ray wavelengths, is associated with particle acceleration
within or near the white dwarf magnetosphere \cite{Geng2016,Takata2018}. Figure 4 shows the spectral energy distibutions of AR Sco.
Thermal components from the white dwarf and M star are also present \cite{marsh2016}, while
XMM-Newton observations show persistent optically thin thermal plasma emission as well as a pulsed non-thermal component \cite{Takata2021}.

\begin{figure}[t]
\centering
\includegraphics[width=.92\textwidth]{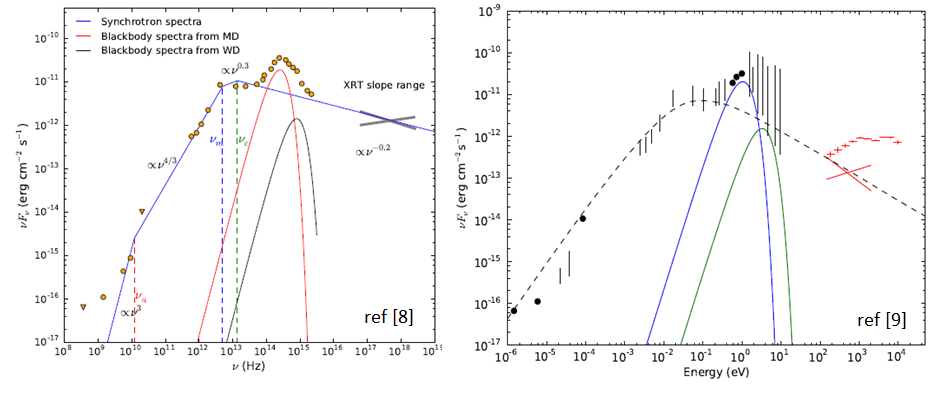}
\caption{Broadband spectral energy distribution of AR Sco (\cite{Geng2016,Takata2018}). The system is
    remarkable because non-thermal emission dominates over much of the
    electromagnetic spectrum, unlike those of accretion-powered magnetic cataclysmic
    variables. In the right panel \cite{Takata2018}, the XMM-Newton X-ray component is shown in red.}
\label{fig:arsco_sed}
\end{figure}

\subsection{Spin-down power and the magnetic dipole interpretation}
\label{subsec:arsco_spindown}

The observed spin period increase discovered in \cite{marsh2016} (later revised in \cite{pelisoli2022}; see Figure 5) implies that the compact object is losing
rotational kinetic energy. If the compact object is treated as a neutron
star, the available spin-down luminosity is only about
$1.1\times10^{28}\,{\rm erg\,s^{-1}}$. For a white dwarf, however, the much
larger moment of inertia gives a spin-down power of order
$1.5\times10^{33}\,{\rm erg\,s^{-1}}$. The observed pulsed luminosity of the
system, $L_{\rm pulse}\simeq0.6$--$3.6\times10^{32}\,{\rm erg\,s^{-1}}$, can
therefore be supplied by the rotational energy loss of a white dwarf, but not
by a neutron star.

This argument is one of the clearest reasons why AR Sco cannot be understood
as a hidden neutron-star system. The low X-ray luminosity, low
$L_{\rm X}/L_{\rm opt}$ ratio, and distance constraints are also inconsistent
with a conventional neutron-star binary. The natural interpretation is
therefore magnetic dipole spin-down of a rapidly rotating white dwarf.

\begin{figure}[t]
\centering
\includegraphics[width=.92\textwidth]{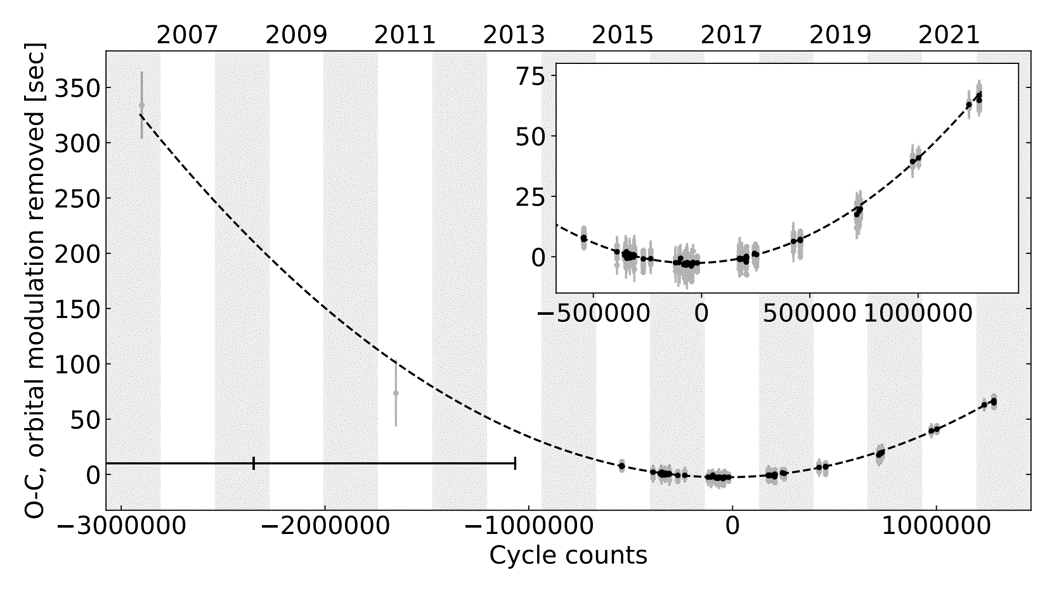}
\caption{The spin-down of AR Sco \cite{pelisoli2022}. }
\label{fig:arsco_spindown}
\end{figure}

\subsection{Why AR Sco is not an ordinary magnetic cataclysmic variable}
\label{subsec:arsco_not_cv}

An obvious question following the discovery was whether AR Sco could simply
be an unusual cataclysmic variable. Several observational facts argue against this
interpretation. The optical light curves do not show any stochastic
flickering expected from active accretion. The optical spectra do not show
the broad or complex emission-line profiles associated with an accretion
disc, stream, or accretion curtains. The X-ray luminosity is only
$L_{\rm X}\sim5\times10^{30}\,{\rm erg\,s^{-1}}$, roughly two orders of
magnitude below that of many magnetic cataclysmic variables.

The system is also different in its polarisation behaviour \cite{buckley2017}, discussed in
detail in Section~\ref{sec:polarimetry}. Accreting magnetic white dwarfs
typically show cyclotron-dominated emission, often with strong circular
polarisation (up to tens of \%). AR Sco instead shows exceptionally strong linear polarisation (40\%)
and low circular polarisation (few \%) \cite{PotterBuckley2018}. This distinction is a central
reason why AR Sco is better understood as a rotationally powered white dwarf
pulsar rather than as a member of the established magnetic-CV class.

\subsection{Summary of the AR Sco characteristics}
\label{subsec:arsco_summary}

The defining observational facts for AR Sco can be summarised as follows:
it shows coherent spin and beat pulsations; non-thermal emission dominates
the spectrum; the white dwarf is spinning down; the spin-down power is
sufficient to power the observed pulsed luminosity; and the system lacks the
usual signatures of accretion-dominated magnetic cataclysmic variables.
Together these properties established AR Sco as the first compelling example
of a white dwarf pulsar.

The timing, spectral energy distribution, and energetics already show that
AR Sco is powered primarily by rotational energy loss. However, they do not
by themselves identify the detailed emission geometry. That evidence came
from high-speed optical polarimetry, which revealed record-breaking linear
polarisation and a rotating-vector behaviour strikingly reminiscent of
neutron-star pulsars.

\section{Polarimetry: the defining evidence for a rotating dipole}
\label{sec:polarimetry}

\begin{figure}
\centering
\includegraphics[width=1.0\textwidth]{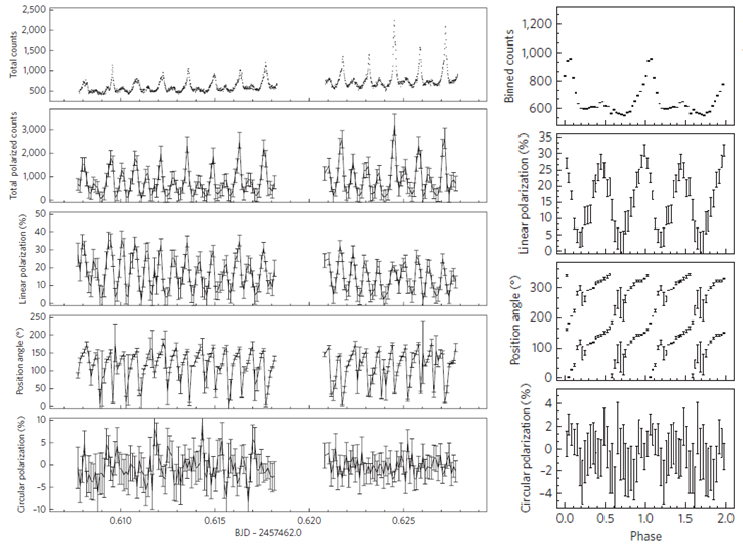}
\caption{First high-speed HIPPO observations of AR Sco taken on 14 March 2016 (left panel) showing strong pulsed linear
polarisation plus rotation of the linear polarisation vector through 180$^{\circ}$. The right panel shows the phase-folded variations on the beat period \cite{buckley2017}.} 
\label{fig:hippo_first}
\end{figure}

\subsection{The first HIPPO observations}

The discovery of AR Sco immediately suggested that polarimetry might provide
important clues to the emission mechanism. Initial all-Stokes observations
were obtained on 14 and 15 March 2016 with the HIPPO high-speed
photo-polarimeter on the SAAO 1.9-m telescope \cite{buckley2017}. In Figure 6 the polarimetry is shown for the 14 March observation. Even during the observations
it was apparent from the real-time data that the system exhibited
extraordinarily strong and rapidly varying polarisation. Subsequent reduction
showed that the modulation occurred on the $\sim2$ min spin/beat cycle,
leading directly to the interpretation that the polarised emission was linked
to the rotating white dwarf magnetosphere.

\subsection{Linear and circular polarisation}

The most remarkable observational result was the strength of the linear
polarisation. It varies from nearly 0 to almost 40\%, with pulse fractions
approaching 90\% (see Figure 6). At the same time, very little circular polarisation is
present. This behaviour is fundamentally different from that of accreting
magnetic cataclysmic variables, where cyclotron emission typically produces
substantial circular polarisation.

The electric-vector position angle also rotates through
180$^\circ$ during each spin cycle, consistent with emission from a rotating
magnetic dipole. These observations immediately suggested that synchrotron,
rather than cyclotron, emission dominates the optical pulsations.

\begin{figure}[t]
\centering
\includegraphics[width=.92\textwidth]{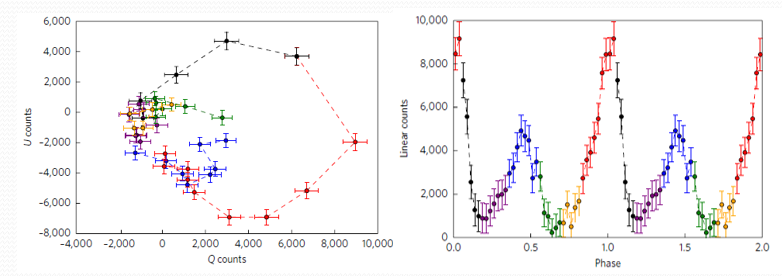}
\caption{Variations of Stokes $Q$ and $U$ on one night over the spin cycle (left), colour-coded to the points shown in the phase-folded light curve (right). The looping behaviour is reminiscent of the Crab pulsar  \cite{buckley2017}.}
\label{fig:linpol}
\end{figure}

\subsection{Spin-folded behaviour}

Phase-folding on the spin period reveals that the polarisation profile
changes with orbital phase. The interaction between the spin and orbital
motions produces different viewing aspects of the irradiated companion and
the emitting regions. Consequently both the polarisation amplitude and
waveform depend on orbital phase.

In the Stokes $Q-U$ plane the polarisation traces characteristic loops during each
spin cycle (see Figure 7). Their morphology closely resembles that seen in optical
polarimetry of the Crab pulsar and is naturally interpreted in terms of a rotating magnetosphere.

\subsection{Extended observing campaigns}

Motivated by the discovery observations, extensive observing campaigns were
undertaken during 2016 and 2017, accumulating more than 65 hours of
high-speed photo-polarimetry over many orbital cycles \cite{PotterBuckley2018}. The resulting power
spectra (e.g. Figure \ref{fig:arsco_powerspectrum}) show that the apparent complexity can be explained by only two
fundamental frequencies, namely the orbital and spin periods, together with
their sidebands and observational aliases.

These data also enabled the construction of a geometrical model \cite{PotterBuckley2018} (see Figure 8) in which the
polarised synchrotron emission originates predominantly from two magnetic
poles, with the observed modulation of the polarization depending on orbital aspect and beaming.

\begin{figure}[t]
\centering
\includegraphics[width=.75\textwidth]{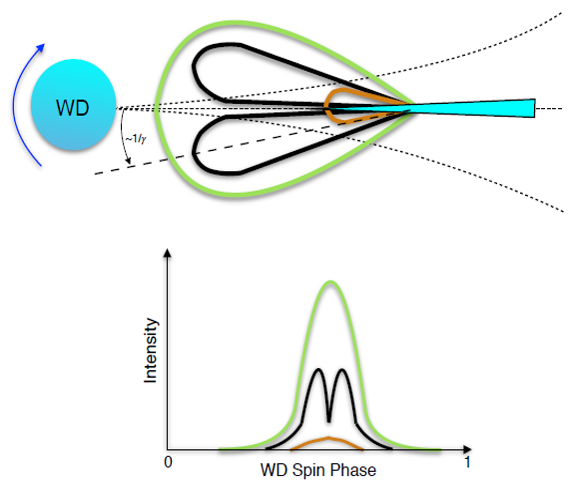}
\caption{Illustrative geometrical model for the polarised emission from
AR Sco. Linear polarisation pulses (black) have a double-lobed intensity profile, while circular (brown) and total intensity (green) are single-lobed. }
\label{fig:geometry}
\end{figure}

\subsection{Rotating vector model}

The polarisation position-angle swing naturally motivated application of the
rotating vector model (RVM), originally developed for radio pulsars.
Subsequent modelling \cite{duPlessis2022,duPlessis2024} demonstrated that the observed position-angle variations
can be reproduced by a highly inclined magnetic dipole. MCMC fitting yielded
constraints on the magnetic obliquity and inclination and supported a white
dwarf mass close to $1\,M_\odot$. The modelling also favours synchrotron
radiation from relativistic particles with relatively small pitch angles.

\subsection{Summary of AR Sco}

The polarimetric observations transformed AR Sco from an unusual interacting
binary into the first convincing example of a white dwarf pulsar. The
combination of record-breaking linear polarisation, weak circular
polarisation, a 180$^\circ$ rotation of the polarisation position angle, and
Crab-like behaviour in the Stokes parameters provides the strongest evidence
for a pulsar-like magnetosphere surrounding the white dwarf.

\section{Multiwavelength campaigns}
\label{sec:multiwave}


Following the discovery and polarimetric confirmation of AR Sco as a
rotationally powered white dwarf, coordinated multiwavelength campaigns were
undertaken to investigate the system across the electromagnetic spectrum.
These campaigns combined optical photometry, spectroscopy, X-ray and radio
observations obtained simultaneously wherever possible.

\subsection{The June 2020 campaign}

\begin{figure}[t]
\centering
\includegraphics[width=.80\textwidth]{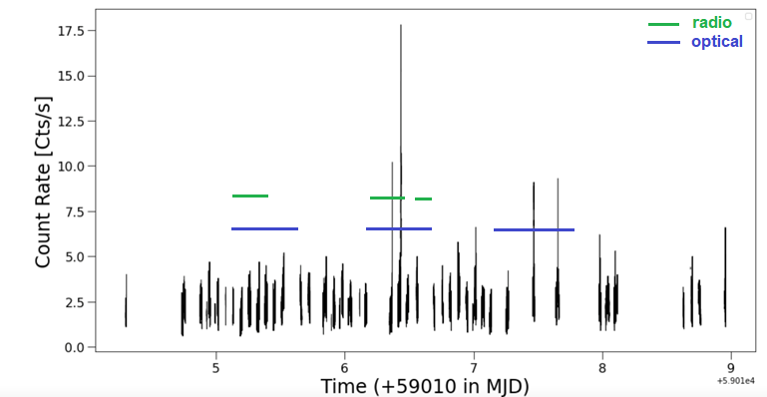}
\caption{The timelines for the June 2020 campaign observations. Black points are the NICER windows, blue are the SAAO/SALT photometric/spectroscopic observations and green the MeerKAT DDT radio observations.}
\label{fig:2020}
\end{figure}

A major campaign was conducted during 14--16 June 2020. The principal optical
component consisted of time-resolved spectroscopy with SALT together with half a night of
Keck spectroscopy. Simultaneous high-speed optical photometry was obtained
with SHOC on the SAAO 1-m telescope. By fortunate coincidence, NICER X-ray
observations (PI: K. Mori) and MeerKAT Director's Discretionary Time radio
observations overlapped much of the campaign. Results from this campaign will appear in forthcoming papers. In Figure \ref{fig:2020} the timelines for the different observations during this campaign are presented. 

\subsection{Time-resolved SALT spectroscopy}

\begin{figure}[t]
\centering
\includegraphics[width=1\textwidth]{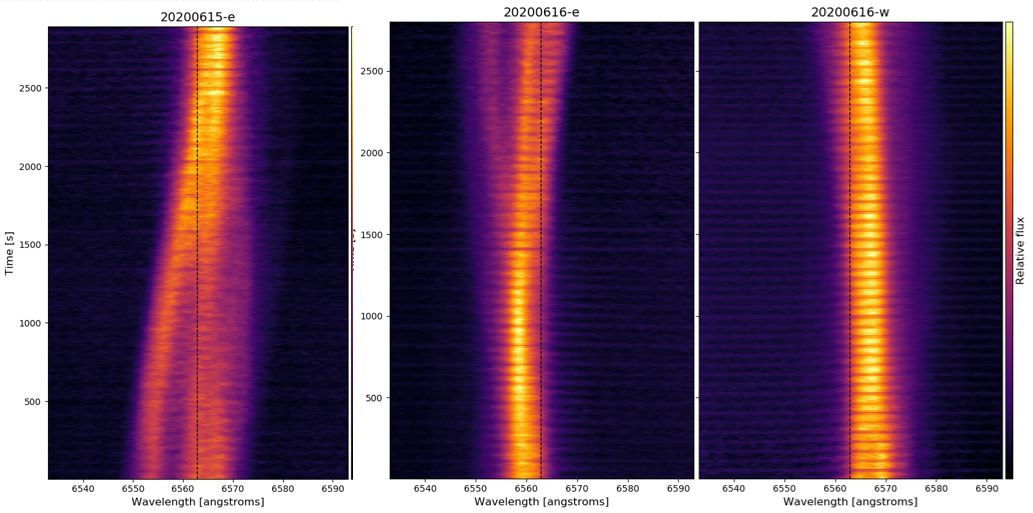}
\caption{Three SALT tracks showing the trailed spectra of the H$\alpha$ line, taken with 10-s exposures. Intensity pulsations are seen in both the lines and continuum at the first and second harmonics of the beat period. }
\label{fig:trailed-spectra}
\end{figure}

Six $\sim$3000-s-long SALT tracks were obtained over three nights, sampling essentially the complete
3.56-h orbital cycle. Using frame-transfer mode with 10-s integrations and no
dead-time, the spectra revealed pulsations in both the continuum and the
emission lines. The pulse amplitudes and phases vary with wavelength,
providing important constraints on the emitting regions.

Trailed spectra (Figure 10) and Doppler tomography (Figure 11) show that much of the Balmer emission
originates on the trailing hemisphere of the irradiated secondary star. Spin
tomography further suggests emission associated with both magnetic poles.

\begin{figure}[t]
\centering
\includegraphics[width=0.5\textwidth]{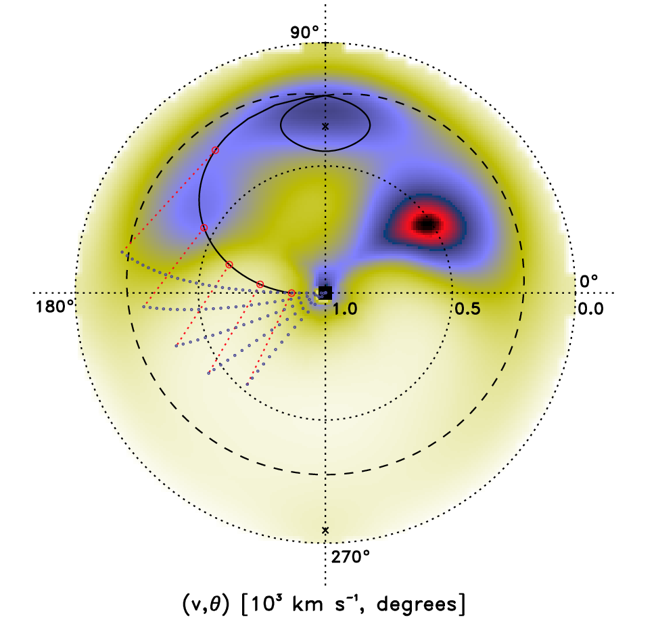}
\caption{Inside-out Doppler tomogram of the H$\alpha$ line from the combined SALT spectra. The bulk of the emission arises in the trailing region of the secondary star.}
\label{fig:tomo}
\end{figure}

\subsection{NICER X-ray observations}

NICER detected coherent pulsations associated with the white dwarf. Power
spectra and folded light curves demonstrate that the X-ray emission is linked
to the same rotating magnetosphere responsible for the optical pulsations,
although the detailed pulse morphology differs with energy.

\subsection{MeerKAT radio observations}

MeerKAT L-band observations clearly detected AR Sco as a variable radio
source. Simultaneous optical observations enabled direct comparison between
radio and optical pulse behaviour. At L-band the radio emission appears more
closely associated with the white dwarf spin period than the optical beat
period. Variable linear polarisation is detected together with weaker
circular polarisation.

The radio observations reveal behaviour that differs from earlier VLA
results, suggesting frequency-dependent emission properties and providing new
constraints on the radio-emitting particle population.

\subsection{Implications}

The combined optical, X-ray and radio observations demonstrate that no single
waveband provides a complete description of AR Sco. Instead, the pulsar-like
behaviour is expressed differently across the spectrum, reflecting particle
acceleration, synchrotron emission and the interaction of the rotating white
dwarf magnetosphere with its companion.

These coordinated campaigns provide the observational foundation for current
models of the system and motivate similar studies of newly discovered white
dwarf pulsars.

\section{The expanding family of white dwarf pulsars}
\label{sec:newsystems}

\subsection{Searching for AR Sco analogues}

Following the discovery of AR Sco, extensive searches were undertaken to find
similar systems. Candidate selection using eROSITA, Gaia and WISE finally led
to the identification of a second convincing white dwarf pulsar in 2022,
eRASSU J191213.9$-$441044 (J1912), independently by two groups \cite{pelisoli2023,Schwope2023}.

\subsection{J1912-4410}

J1912 has an orbital period of about 4.03 h and a white dwarf spin period of
320 s. Optical photometry reveals coherent pulsations together with
flickering and flaring that may indicate a greater contribution from mass
transfer than in AR Sco.

Time-resolved SALT spectroscopy confirms pulsations at the spin period, while
HIPPO polarimetry shows strong pulsed linear polarisation with behaviour
closely resembling AR Sco, although dominated by a single magnetic pole.
MeerKAT observations detected narrow pulsed radio emission, but only at certain orbital phases, while XMM-Newton
observations showed both thermal plasma and power-law X-ray components, plus evidence for flares.

\subsection{The third system: SDSS J230641.47+244055.8}

A third member of the class, SDSS J230641.47+244055.8 (J2306), was recently
identified \cite{castro2025}. Initially classified as a contact binary and later as a
cataclysmic variable, high-speed ULTRACAM photometry revealed coherent
92-second pulsations, establishing its AR Sco-like nature.

The optical spectrum again shows a blue continuum together with an M-star
companion and narrow Balmer emission lines, making it closely resemble both
AR Sco and J1912.

\subsection{Comparison of the three systems}

Although AR Sco, J1912 and J2306 clearly belong to the same class, important
differences are emerging. J1912 shows evidence for optical flickering, unlike AR Sco. The radio spin pulses are also very intense and narrow, occurring only at certain orbital phases, whereas AR Sco displays more variable radio behaviour and its pulses are revealed only in the power spectra. These
differences probably reflect variations in magnetic geometry, spin period,
mass-loss rate and evolutionary state.

As the sample increases it will become possible to determine which observed
properties are fundamental characteristics of white dwarf pulsars and which
are specific to individual systems.

\subsection{Outlook}

The discovery of three systems within a few years demonstrates that AR Sco is
not unique. Future surveys, particularly those combining X-ray, optical and
time-domain information, are expected to reveal additional examples and allow
the population properties of white dwarf pulsars to be explored.

\section{Evolutionary implications}
\label{sec:evolution}

The discovery of three confirmed white dwarf pulsars raises the question of
their evolutionary origin. These systems require the coexistence of a rapidly
rotating white dwarf and a strong magnetic field, a combination that is not
easy to produce within standard models of magnetic cataclysmic variables.

A leading evolutionary scenario proposes that the white dwarf was spun up
during an earlier phase of high mass transfer. The magnetic field may have
been temporarily buried by accreted material and subsequently re-emerged
after accretion declined. In this picture the system evolves through a brief
rotationally powered phase before eventually becoming synchronised as a
polar.

In \cite{Schreiber2021}, it is proposed that AR Sco-like systems represent a short
transitionary stage linking intermediate polars and polars, perhaps with a very short timescale of $\sim10^7$ years. Alternative ideas
include magnetic-field evolution associated with crystallisation or diffusion
processes within the white dwarf interior. These possibilities remain under
active investigation.

Several observational issues remain unresolved. It is not yet clear why only
a handful of systems are currently known, whether all such systems pass
through a white dwarf pulsar phase, or what controls the differing behaviour
seen in AR Sco, J1912 and J2306. Continued searches for additional systems
will be essential for distinguishing between competing evolutionary models.

Future progress will depend on coordinated time-domain surveys, sensitive
radio facilities such as MeerKAT and the SKA, continued X-ray monitoring, and
high-speed optical photometry and polarimetry. Increasing the sample size is
likely to provide the strongest constraints on the evolutionary pathways
leading to white dwarf pulsars.

\section{Summary}
\label{sec:summary}

AR Sco established that a white dwarf can power pulsar-like emission through
the loss of rotational kinetic energy. The combination of coherent spin and
beat pulsations, a spectral energy distribution dominated by non-thermal
emission, measurable spin-down, and remarkable optical polarimetric
properties demonstrates that the system is fundamentally different from
accretion-powered magnetic cataclysmic variables.

Subsequent multiwavelength observations have strengthened this picture.
Optical spectroscopy, X-ray timing and radio observations provide a coherent
view of particle acceleration and synchrotron emission within a rotating
white dwarf magnetosphere interacting with its companion star. In addition, the recent discovery of rotational pulsations in the Fermi-LAT data of AR Sco strengthens the claim of a pulsar-like particle acceleration mechanism is at work \cite{Kaplan2022}  

The discoveries of J1912 and J2306 demonstrate that AR Sco is not a
unique object but the prototype of an emerging class of white dwarf pulsars.
At the same time, differences between the known systems suggest that a range
of magnetic geometries, mass-transfer/loss rates and evolutionary states may be
represented.

The most compelling observational evidence continues to come from optical
polarimetry. The exceptionally strong linear polarisation, weak circular
polarisation, rotating position angle and Crab-like behaviour of the Stokes
parameters provide persuasive evidence for a pulsar-like magnetosphere.

Many important questions remain concerning the origin, evolution and emission
physics of these systems. Nevertheless, the discovery of white dwarf pulsars
has opened a new area of compact-object astrophysics, demonstrating that
pulsar-like behaviour is not unique to neutron stars.

Finally, with the advent of the Rubin Observatory's Legacy Survey of Space and Time (LSST), there are also prospects to discover more such systems from their light curves and colours, notwithstanding the few day cadence. 

\end{document}